# Comment on "Linear wave dynamics explains observations attributed to dark-solitons in a polariton quantum fluid"

A. Amo[1], J. Bloch[1], A. Bramati[2], I. Carusotto[3], C. Ciuti[4], B. Deveaud-Plédran[5], E. Giacobino[2], G. Grosso[5], A. Kamchatnov[6], G. Malpuech[7], N. Pavloff[8], S. Pigeon[9], D. Sanvitto[10], D. D. Solnyshkov[7]

[1]*Laboratoire Photonique et Nanostructures, CNRS, Route de Nozay, 91460, Marcoussis, France*
[2]*Laboratoire Kastler Brossel, Université Pierre et Marie Curie, Ecole Normale Supérieure et CNRS, UPMC Case 74, 4 place Jussieu, 75252 Paris Cedex 05, France*
[3]*INO-CNR BEC Center and Università di Trento, 38123 Povo, Italy*
[4]*Laboratoire Matériaux et Phénomènes Quantiques, Université Paris Diderot-Paris 7 et CNRS, Bâtiment Condorcet, 10 rue Alice Domon et Léonie Duquet, 75205 Paris Cedex 13, France*
[5]*Institute of Condensed Matter Physics, École Polytechnique Fédérale de Lausanne (EPFL), CH-1015 Lausanne, Switzerland*
[6]*Institute of Spectroscopy, Russian Academy of Sciences, Moscow, Troitsk, 142190, Russia*
[7]*Institut Pascal, PHOTON-N2, Clermont Université, Blaise Pascal University, CNRS, 24 Avenue des Landais, 63177 Aubière Cedex, France*
[8]*Università Paris Sud, CNRS, Laboratoire de Physique Théorique et Modèles Statistiques, UMR8626, F-91405 Orsay, France*
[9]*School of Mathematics and Physics, Queen's University Belfast, BT7 1NN, Northern Ireland, UK*
[10]*NNL, Istituto Nanoscienze - Cnr, Via Arnesano, 73100 Lecce, Italy*

**In a recent preprint (arXiv:1401.1128v1) Cilibrizzi and co-workers report experiments and simulations showing the scattering of polaritons against a localised obstacle in a semiconductor microcavity. The authors observe in the linear excitation regime the formation of density and phase patterns reminiscent of those expected in the non-linear regime from the nucleation of dark solitons. Based on this observation, they conclude that previous theoretical and experimental reports on dark solitons in a polariton system should be revised. Here we comment why the results from Cilibrizzi et al. take place in a very different regime than previous investigations on dark soliton nucleation and do not reproduce all the signatures of its rich nonlinear phenomenology. First of all, Cilibrizzi et al. consider a particular type of radial excitation that strongly determines the observed patterns, while in previous reports the excitation has a plane-wave profile. Most importantly, the nonlinear relation between phase jump, soliton width and fluid velocity, and the existence of a critical velocity with the time-dependent formation of vortex-antivortex pairs are absent in the linear regime. In previous reports about dark soliton and half-dark soliton nucleation in a polariton fluid, the distinctive dark soliton physics is supported both by theory (analytical and numerical) and experiments (both continuous wave and pulsed excitation).**

The non-linear Schrödinger (Gross-Pitaevskii) equation describing a coherent field with a positive effective mass and repulsive interparticle interactions allows non-linear localised solutions known as dark solitons [1]. In two-dimensions, it was recently shown theoretically that dark solitons can nucleate and stabilise in the wake of a suitably large and strong obstacle present in the flowpath of a fluid [1, 2]. The same physics describes the case of polariton propagation with losses [3]. In this situation, dark solitons present at least the following properties, which allow their identification:

1. They appear as oblique dark intensity lines in the wake of the obstacle [1].
2. There is an abrupt jump in the phase of the field across the soliton. The magnitude of the phase jump is between 0 and π.
3. The phase jump, the soliton width and depth, and the fluid velocity and density are related by a very precise analytical expression that can be deduced from the above mentioned non-linear Schrödinger equation [4-6].

4. There exists an upper critical density for the nucleation of stable solitons. This critical density depends on the flow velocity [1, 2].
5. Above the critical density, nucleation of solitons is replaced by other kinds of behaviour, such as a time-dependent regime with an (almost) periodic emission of vortices and antivortices and, at even larger densities, a superfluid behaviour [1, 7, 8].

Even though solitons are intrinsically non-linear entities, some of their typical features can also be observed in strictly linear systems. An example is reported in the recent publication by Cilibrizzi and collaborators [9]. In their work, Cilibrizzi et al., show experiments and simulations in which a non-interacting polariton fluid encountering an obstacle in a planar microcavity shows some features recalling those of dark solitons. The obstacle is a spatially extended photonic trap whose attractive potential is able to deflect and diffract the incident polariton field. The interference of the incident and scattered waves then gives rise to spatial features in the wake of the obstacle. In particular, these features show properties 1 and 2 mentioned above: oblique dark intensity lines and a phase jump across them. The authors also show that when changing the frequency and wavevector of the incident polariton field, the dark lines disappear. In their case, the disappearance phenomenon is related to the fine interplay between the incident wavevector and the phase shift while crossing the obstacle, needed to generate the observed features.

From their observations in the linear regime, Cilibrizzi and coworkers conclude that: "Therefore, the previous reports of the observation of dark-solitons [8, 10-13] and half-dark-solitons [14, 15] which were based on these features have to be reconsidered."

Unfortunately Cilibrizzi et al. concentrate their attention on the dark soliton images only and do not take in due consideration the numerous other observations that were presented in Refs. [8, 10-15]. The work reported in these references shows that all 5 conditions stated above are satisfied: it was on the basis of this rich experimental evidence that the observations were carefully attributed to dark solitons and half-dark solitons. In particular, all the theoretical and experimental works in those references are performed in the non-linear regime, as attested by the significant power dependence explicitly shown, for instance, in Fig. 1 of [8], Figs. 3 and S3 of [10], and Fig. 3 of [11]. On the other hand, Cilibrizzi and co-workers state that they can mimic the power dependence by changing the wavevector and energy of the incident polaritons, the argument being that an increase in interactions in the polariton gas induces a blueshift and therefore an acceleration. However in the soliton experiments in Refs. [8, 10-15] the trend is in the opposite direction: (*i*) the total energy of the injected polaritons is imposed by the laser and kept constant when changing the power, (*ii*) since interaction energy plus kinetic energy is constant and fixed by the laser, the increase of the interaction energy due to a more intense pump must be compensated by a corresponding decrease of the kinetic energy. Thus the net effect of interactions on the speed of polaritons is to slow them down, in stark contrast to Cilibrizzi et al.'s argument. Furthermore, Cilibrizzi et al. fail to show any evidence of the peculiar time-dependent regime characteristic of the nonlinear Schrödinger equation [7, 8] with a (almost) periodic nucleation of vortex-antivortex pairs, which was instead observed in [16, 17] and [10] when the pump intensity is tuned in between the soliton and the superfluid regimes.

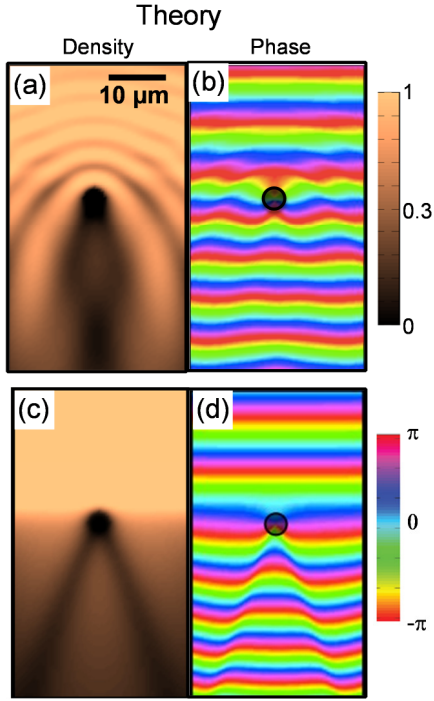
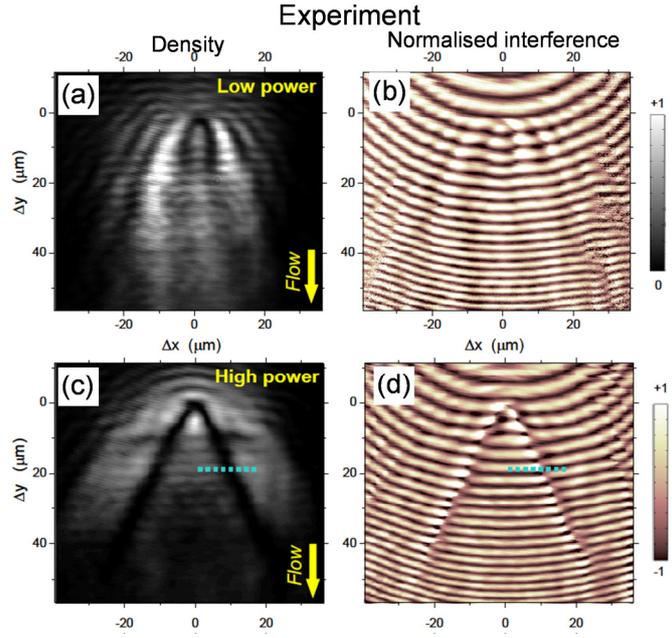

FIG. 1. Simulations based on a Schrödinger equation with pump and losses in the linear (upper panels, (a)-(b)) and nonlinear (lower panels, (c)-(d)) regimes. In both simulations, polaritons are injected above the obstacle with a single wavevector value pointing downwards and only the density is varied. (c)-(d) are reproduced from Fig. 1 in Ref. [8], while (a)-(b) were calculated in the same conditions without non-linearity.

FIG. 2.- Experimental images of the intensity and phase of a polariton fluid in a semiconductor microcavity in the low power (linear) regime showing no solitons (a)-(b), and in the high power (non-linear) regime showing the nucleation of two oblique dark solitons (c)-(d). Polaritons are injected above the obstacle with a single wavevector corresponding to a flow in the downward direction. All parameters are kept constant except for the incident power. From the Supplementary Online Material in Ref. [10]. Note that the curved shape of the interference maxima in (b) and (d) is an experimental artefact related to the geometry of the reference beam in the interferometric imaging.

An additional difference between the work from Cilibrizzi et al. and those in Refs. [8, 10-15], not discussed by Cilibrizzi and coworkers, is that to observe the dark traces and the phase jumps in the wake of the obstacle the *authors use a fluid with a radial expansion* (see supplementary information in [9]). This is a very different experimental scheme from the one used in Refs. [8, 10-15] and from similar works in the context of atomic physics [1, 18], where the quantum fluid is generated in the shape of a plane-wave with a single well-defined momentum. A simulation performed with an incident field in a plane-wave form in the linear regime is shown in Fig. 1(a)-(b). No soliton-like traces are present in the wake of the obstacle. This result is similar to the experiment reported in Fig. S3 of Ref. [10] and shown in Fig. 2(a)-(b), obtained at low power in the linear regime with a plane-wave-like pump. Keeping all parameters equal but increasing the polariton density high enough to enter the non-linear regime, both simulations and experiments (Fig. 1(c)-(d) and Fig. 2(c)-(d)) show the formation of oblique dark soliton pairs.

In conclusion, Cilibrizzi and coworkers have provided experimental evidence of interference and diffraction effects of polaritons (or photons) hitting an obstacle in the linear regime. In particular, their results have shown that the observation of dark traces accompanied by sudden phase jumps of the field are not exclusive of soliton physics, but they can also appear in the linear regime of a non-interacting field. However, contrary to what is stated in Cilibrizzi et al.'s work, the results published in Refs. [8,

10-15] do indeed address dark soliton physics because they experimentally demonstrate every single criterion stated above from 1 to 5, and, in particular, they show the crucial role of nonlinear effects in determining the critical densities separating the formation of stable solitons from other regimes showing, e.g., a superfluid behaviour or a time-dependent nucleation of vortices and anti-vortices. To the best of our knowledge, all experiments in Refs. [8, 10-15] can only be reproduced with a Schrödinger equation model including non-linearities.